\documentclass[11pt]{article}
\usepackage{amssymb, amsmath, stmaryrd}
\usepackage{graphicx}
\usepackage{bbm}
\newcommand{\beq}[1]{\begin{equation}\label{#1}}
\newcommand{\eeq}{\end{equation}}
\newcommand{\req}[1]{(\ref{#1})}
\newcommand{\bmu}[1]{\begin{multline}\label{#1}}
\newcommand{\emu}{\end{multline}}

\renewcommand{\(}{\left(}
\renewcommand{\)}{\right)}
\renewcommand{\[}{\left[}
\renewcommand{\]}{\right]}

\newcommand{\eq}{\triangleq}

\newcommand{\A}{\mathcal{A}}

\renewcommand{\S}{{\mathcal S}}

\newcommand{\x}{{\textbf{\textit{x}}}}

\renewcommand{\u}{{\textbf{\textit{u}}}}
\renewcommand{\v}{{\textbf{\textit{v}}}}

\begin{document}

\begin{center}
{\Large\bf Threshold Disjunctive Codes}
\\[15pt]
{\bf A.G. D'yachkov, \quad I.V. Vorobyev, \quad N.A. Polyanskii,\quad V.Yu. Shchukin}
\\[15pt]
Moscow State University, Faculty of Mechanics and Mathematics,\\
Department of Probability Theory, Moscow, 119992, Russia,\\
{\sf agd-msu@yandex.ru,\quad vorobyev.i.v@yandex.ru,\quad nikitapolyansky@gmail.com,\quad vpike@mail.ru}
\end{center}

\medskip

{\bf Abstract.}\quad
Let $1 \le s < t$, $N \ge 1$ be integers and a complex electronic
circuit of size $t$ is said to be an $s$-active, $\; s \ll t$,
and can work as a system block if
not more than $s$ elements of the circuit are defective.
Otherwise, the circuit is said to be an $s$-defective and should be substituted
for the $s$-active circuit.
Suppose that there exists a possibility to check the $s$-activity of the circuit
using $N$ non-adaptive group tests identified by a conventional
disjunctive $s$-code $X$ of size~$t$ and length~$N$.
As usually, we say that any group test yields the positive response
if the group contains at least one defective element.
In this case, there is no any interest to look for
the defective elements. We are keen to decide on the number of
the defective elements in the circuit without knowing the code~$X$.
In addition, the decision has the minimal possible complexity because it
is based on the simple comparison of a fixed threshold $T$, $0 \le T \le N - 1$,
with the number of positive responses $p$, $0 \le p \le N$,
obtained after carrying out $N$ non-adaptive tests prescribed by the disjunctive $s$-code~$X$.
For the introduced group testing problem, a new class of the well-known
disjunctive $s$-codes called the threshold disjunctive $s$-codes is defined.
The aim of our paper is to
discuss both some constructions of suboptimal threshold disjunctive $s$-codes and
the best random coding bounds on the rate of threshold disjunctive $s$-codes.

\section{Notations, Definitions and Statement of Problems}
\indent \indent
Let $N$, $t$, $s$ and $T$ be integers, where $2 \le s < t$ and $0 < T < N$.
Let $\eq$ denote the equality by definition, $|A|$ -- the size of the set $A$ and
$[N] \eq \{1, 2, \dots, N\}$ -- the set of integers from $1$ to~$N$. The standard symbol
$\lfloor a \rfloor$ will be used to denote the largest integer~$\le a$.

A binary $(N \times t)$-matrix
$$
X = \| x_i(j) \|, \quad x_i(j) = 0, 1, \quad \x_i \eq (x_i(1), \dots, x_i(t)), \quad \x(j) \eq (x_1(j), \dots, x_N(j)),
$$
$ i \in [N] $, $ j \in [t] $, with $N$ rows $\x_1, \dots, \x_N$ and $t$ columns $\x(1), \dots, \x(t)$ (codewords)
is called a \textit{binary code of length $N$ and size $t = \lfloor 2^{RN} \rfloor$},
where a fixed parameter $R > 0$ is called a \textit{rate} of the code~$X$.
The number of $1$'s in the codeword $x(j)$, i.e.,
\beq{w}
|\x(j)| \eq \sum\limits_{i = 1}^N \, x_i(j), \qquad j \in [t],
\eeq
is called the \textit{weight} of $x(j)$, $j \in [t]$. A code $X$ is called
a \textit{constant weight binary code of weight $w$}, $1 \le w < N$, if for any $j \in [t]$, the weight $|\x(j)| = w$.

Let the conventional symbol $\u \bigvee \v$ denote the disjunctive (Boolean) sum of binary columns $\u, \v \in \{0, 1\}^N$.
We say that a column $\u$ \textit{covers}
a column $\v$ ($\u \succeq \v$) if $\u \bigvee \v = \u$.

\textbf{Definition 1.}~\cite{dr83,ks64}.\quad
A binary code $X$ is said to be a \textit{disjunctive $s$-code}
if the disjunctive sum of any $s$-subset
of codewords of $X$ covers those and only those
codewords of $X$ which are the terms of the given disjunctive sum.

Let $\S$, $\S \subset [t]$, $ |\S| \le s$, be an arbitrary fixed collection
of defective elements of size~$|\S|$ and the corresponding binary \textit{response vector} is defined as follows:
\beq{rv}
\x(\S) \eq \bigvee_{j \in \S} \, \x(j), \qquad \S \subset [t], \quad |\S| \le s.
\eeq
For the classical problem of non-adaptive group testing, let us depict $N$ tests
as an binary $(N \times t)$-matrix $X = \| x_i(j) \|$,
where a column $\x(j)$ corresponds to the $j$-th element, a row $\x_i$ corresponds to the $i$-th test and
$x_i(j) \eq 1$ if and only if the $j$-th element is included into the $i$-th testing group.
The result of a test equals $1$ if at least one defective element is included into the testing group
and $0$ otherwise, so the column of results is exactly $\x(\S)$.
The definition of disjunctive $s$-code $X$
gives the important sufficient condition for the evident identification of any such collection $\S$.
In this case the identification is to find all codewords of the code~$X$ covered by the response vector. Thus,
it significantly depends on
the code $X$ and its complexity equals the code size~$t$.

Let $T$, $0 < T < N$, be an arbitrary integer parameter.
One can easily understand that for any such parameter $T$, a sufficient condition for the conventional
disjunctive $s$-code $X$ of length $N$ and size $t$, applying to
the group testing problem described in the abstract of our paper
can be given as

\textbf{Definition 2.}
\quad
A disjunctive $s$-code $X$ of length $N$ and size $t$ is said to be a \textit{threshold disjunctive $s$-code}
with \textit{threshold} $T$ (or, briefly, \textit{disjunctive $s^{T}$-code})
if the disjunctive sum of any $\le s$ codewords of $X$ has weight $\le T$ and
the disjunctive sum of any $\ge s+1$ codewords of $X$ has weight $\ge T + 1$. In other words,
using notations~\req{w} and~\req{rv}, we formally can write that for any collection
of defects $\S$, $\S \subset [t]$, the weight $|\x(\S)|$ of the corresponding
response vector $\x(S)$ satisfies conditions
\beq{def2}
\begin{cases}\quad
|\x(\S)| \; \le T, &\text{if the size $|\S| \le s$},\\
\quad |\x(\S)| \; \ge T+1, &\text{if the size $|\S| \ge s+1$}.
\end{cases}
\eeq
\medskip

\textbf{Remark 1.} The concept of threshold disjunctive $s^T$-codes was motivated by
the troubleshooting in complex electronic circuits using a non-adaptive identification scheme which was
considered in~\cite{zlm76} under the assumption that $\le s$ elements of a circuit
become defective.

\textbf{Remark 2.} A similar model of special disjunctive $s$-codes was considered in~\cite{b15},
where the conventional disjunctive $s$-code is supplied with an additional condition:
the weight $|\x(\S)|$ of the response vector of any subset $\S$, $\S \subset [t]$, $|\S| \le s$, is at most $T$.
Note that these codes have a weaker condition that our threshold disjunctive $s$-codes.
In~\cite{b15} authors motivate their group testing model with bounded weight of the response vector
by a risk for the safety of the persons, that perform test, in some contexts,
when the number of positive test results is too big.

Denote by $t(N, s, T)$ the maximal size of disjunctive $s^{T}$-codes of length $N$. To define the rate
of disjunctive of $s^{T}$-codes,
we introduce a parameter $\tau = T / N$, $0 < \tau < 1$, so the \textit{rate} of $s^{\lfloor \tau N \rfloor}$-codes is
\beq{Rstau}
R^\tau(s) \eq \varlimsup_{N \to \infty} \frac{\log_2 t(N, s, \lfloor \tau N \rfloor)}{N}.
\eeq
Note that any value of $\tau$ allows us to find out whether there is at most $s$ defective elements
or there is more than $s$ defective elements. So, we are interesting in the optimal rate
\beq{Rthreshold}
R_{\text{Thr}}(s) \eq \max_{0 < \tau < 1} R^\tau(s).
\eeq

Some suboptimal constructions of threshold disjunctive $s^T$-codes are presented
in Section~2 of our paper. In Section~3, we develop the random coding method based on the ensemble of constant-weight codes
and establish new lower bounds both on the rate of disjunctive $s^{\lfloor \tau N \rfloor}$-codes and on the optimal rate~$R_{\text{Thr}}(s)$.

\section{Constructions of Disjunctive $s^T$- codes} 
\quad
Let
$
\lambda = \max\limits_{u,v} \, \sum\limits_{i=1}^N\, x_i(u) x_i(v)
$
be the \textit{maximal dot product} for codewords of a code~$X$ .
Consider a constant-weight code $X$ of weight $w$, i.e.,
$|\x(u)| = w$, $\; u = 1, 2, \dots, t$, length $N$,
size $t$ and the maximal dot product $\lambda$, $\; 1 \le \lambda < w$.
The following three statements are evident.
\begin{itemize}
\item
The weight of the disjunctive sum of an arbitrary $s$-subset of
codewords is upper bounded by $T \eq ws$.
\item
The weight of the disjunctive sum of an arbitrary $(s+1)$-subset
of codewords is lower bounded by
$$
T' \eq w(s+1)- {s+1 \choose 2} \lambda = w(s+1) - \frac{s(s+1) \lambda}{2}.
$$
\item
If the inequality
$$
T = ws < T' = w(s+1) - \lambda\frac{s(s+1)}{2}
$$
holds, then the corresponding constant-weight code of the weight $w$
will be a threshold disjunctive $s^T$-code with threshold~$T = ws$.
\end{itemize}

This gives the following {\it sufficient condition}
\beq{KS}
w > \lambda \frac{s(s+1)}{2}, \
\eeq
when a constant-weight code of parameters
$w$ and $\lambda$ will be a
threshold disjunctive $s$-code with threshold~$T = ws$. Obviously,
this condition coincides with the well-known
Kautz-Singleton sufficient
condition~\cite{ks64} for the existence of a disjunctive~$(s(s+1)/2)$-code.

Let $\lambda \ge 1$ be an arbitrary integer and
$q \ge \lambda$ be a prime or a prime power.
Kautz-Singleton~\cite{ks64} constructed the family of
constant-weight binary codes which are based on
the Reed-Solomon $q$-nary codes
whose parameters could be
written in the following form
\beq{RS}
t = q^{\lambda + 1},\quad
w = \lambda \left\lfloor \frac{q}{\lambda} \right\rfloor + 1,\quad
N = qw.
\eeq
The family~\req{RS} yields the possibility to construct threshold
disjunctive $s$-codes of strength $s$ with the threshold $T = sw$
where $s$ is the \textit{maximal possible integer satisfying the inequality}~\req{KS}.
Several numerical examples for parameters of the disjunctive $s^T$-codes based
on~\req{KS}-\req{RS} are presented in the Table~\ref{tab1}.

\begin{table}[ht]
\caption{Parameters of the disjunctive $s^T$-codes based on~\req{KS}-\req{RS}}\label{tab1}
\begin{center}
\begin{tabular}{|c|c||c|c|c|c||c|c|c|}
\hline
$q$ & $t$ & $N$ & $w$ & $\lambda$ & $s$ & $T$ & $T'$\\
\hline
\hline
11 & 121 & 132 & 12 & 1 & 4 & 48 & 50\\
\hline
17 & 289 & 306 & 18 & 1 & 5 & 90 & 93\\
\hline
16 & 4096 & 272 & 17 & 2 & 3 & 51 & 56\\
\hline
23 & 12167 & 529 & 23 & 2 & 4 & 92 & 95\\
\hline
32 & 32768 & 1056 & 33 & 2 & 5 & 165 & 168\\
\hline
31 & 923581 & 961 & 31 & 3 & 4 & 124 & 125\\
\hline
16 & 1048576 & 272 & 17 & 4 & 2 & 34 & 39\\
\hline
\end{tabular}
\end{center}
\end{table}

Note that, for an arbitrary threshold disjunctive $s^T$-code from the family~\req{RS}
satisfying the condition~\req{KS}, the following inequality on the rate holds:
$$
R \eq \frac{\log_2 t}{N} = \frac{\log_2 q}{q} \frac{\lambda + 1}{\lambda \left\lfloor \frac{q}{\lambda} \right\rfloor + 1} \le \frac{\log_2 s'}{s'} \frac{2}{s' + 1},
$$
where $s' = \frac{s(s+1)}{2}$. The upper bound in the right-hand side of the previous inequality behaves like
$\frac{16 \log_2 s}{s^4} (1 + o(1))$ as $s \to \infty$.
In Section~3 we provide a lower bound for $R_{\text{Thr}}(s)$, which is significantly better
and has asymptotics $\frac{\log_2 e}{4s^3} (1 + o(1))$ as $s \to \infty$.

\section{Random Coding Bounds}
\quad
\textbf{Theorem 1.}\quad
\textit{For any $s \ge 2$ and $0 \le \tau \le 1$ the inequality}
\beq{lowerRExpr}
R^\tau(s) \ge \underline{R}^\tau(s) \eq \max_{1 - (1 - \tau)^{1 / (s + 1)} < Q < 1 - (1 - \tau)^{1 / s}}
\min \left\{ \frac{\A'(s, Q, \tau)}{s - 1}, \frac{\A(s+1, Q, \tau)}{s} \right\},
\eeq
\textit{holds, where}
\begin{align}
\label{exprA'}
\A'(s, Q, q) &\eq
\begin{cases}
\A(s, Q, q), \quad &\text{ if } Q \le q \le sQ,\\
\infty, \quad &\text{ otherwise},
\end{cases}\\
\label{exprA}
\A(s, Q, q) &\eq (1-q) \log_2(1-q) + q \log_2 \[ \frac{Qy^s}{1-y} \] + sQ \log_2 \frac{1-y}{y} + sh(Q),\\
\label{entropy}
h(Q) &\eq -Q \log_2 Q - (1 - Q) \log_2 [1 - Q],
\end{align}
{\em and $y$ in the right-hand side of~\req{exprA} is the unique root of the equation}
\beq{qSmall}
q = Q \frac{1 - y^s}{1 - y}, \quad 0 < y < 1.
\eeq

\textbf{Proof.}\quad
Fix $s \ge 2$, $0 < \tau < 1$ and a parameter $Q, \; 0 < Q < 1$.
The bound~\req{lowerRExpr}-\req{qSmall} is obtained by the method of random coding over
the ensemble of binary constant-weight codes~\cite{drr89} defined as
the ensemble $E(N, t, Q)$ of binary codes $X$ of length $N$ and size $t$,
where the codewords are chosen independently and equiprobably from the set
consisting of all ${N \choose \lfloor QN \rfloor}$ codewords of a fixed weight $\lfloor QN \rfloor$.

An $s$-subset $\S \subset [t]$ of indices, $|\S| = s$, is called \textit{bad in code $X$} if
the disjunctive sum of codewords of $X$ corresponding indices from $\S$ has weight
greater than $\lfloor \tau N \rfloor$, i.e.
$$
\left| \bigvee_{i \in \S} \x(i) \right| > \lfloor \tau N \rfloor.
$$
An $(s+1)$-subset $\S_{+1} \subset [t]$ of indices, $|\S_{+1}| = s+1$, is called \textit{bad in code $X$} if
the disjunctive sum of codewords of $X$ corresponding indices from $\S_{+1}$ has weight
not greater than $\lfloor \tau N \rfloor$, i.e.
$$
\left| \bigvee_{i \in \S_{+1}} \x(i) \right| <= \lfloor \tau N \rfloor.
$$

A codeword $\x(j)$ is called bad if there exists a bad $s$-subset $\S$, such that $j \in \S$,
or there exists a bad $(s+1)$-subset $\S_{+1}$, such that $j \in \S_{+1}$.
For the ensemble $E(N, t, Q)$, denote the probability of event ``fixed codeword is bad in $X$''
by $P_0(N, t, Q, s, \tau)$, the probability of event ``fixed $s$-subset is bad in $X$'' by
$P_1(N, Q, s, \tau)$ and the probability of event ``fixed $(s+1)$-subset is bad in $X$''
by $P_2(N, Q, s, \tau)$.

Note that the expectation of the number of bad codewords equals
$$
t \cdot P_0(N, t, Q, s, L, \tau),
$$
therefore, if $P_0 \le 1 / 2$, then there exists an $s^\tau$-code $X'$ of length $N$
and size at least $t / 2$. Hence,
\beq{prob0}
\varlimsup_{N \to \infty} P_0(N, 2 \cdot 2^{\lfloor RN \rfloor}, Q, s, L, \tau) < \frac{1}{2}
\eeq
is sufficient condition for the correctness of inequality
\beq{lowerRBase}
R^\tau(s) \ge R.
\eeq
The use of relation
$$
P(N, t, Q, s, L, \tau) \le {t-1 \choose s-1} P_1(N, Q, s, \tau) +
{t-1 \choose s} P_2(N, Q, s, \tau)
$$
yields the following sufficient condition for~\req{prob0} (and for~\req{lowerRBase}):
\beq{prob1}
\varlimsup_{N \to \infty} \[ {2^{\lfloor RN \rfloor + 1}-1 \choose s-1} P_1(N, Q, s, \tau)
+ {2^{\lfloor RN \rfloor + 1}-1 \choose s} P_2(N, Q, s, \tau) \] < \frac{1}{2}.
\eeq
Therefore, one can derive the following sufficient condition for the lower bound~\req{lowerRBase}
\beq{prob1Lim}
R^{\tau}(s) \ge R \quad \text{if} \quad
\begin{cases}
(s-1) R &< \varlimsup_{N \to \infty} \frac{- \log_2 P_1(N, Q, s, \tau)}{N},\\
s R &< \varlimsup_{N \to \infty} \frac{- \log_2 P_2(N, Q, s, \tau)}{N}.
\end{cases}
\eeq

For a fixed $s$-subset $\S \subset [t]$ and integer $k$, introduce event
$$
W_{s, k} \eq \left\{ \left| \bigvee_{j \in \S} \x(j) \right| = k \right\}
$$
To compute the limits in the right-hand side of~\req{prob1Lim}, we represent
probabilities $P_1(N, Q, s, \tau)$ and $P_2(N, Q, s, \tau)$ in the following forms:
\beq{prob1Form}
\begin{split}
P_1(N, Q, s, \tau) &= \sum_{k = \lfloor \max\{\tau, Q\} N \rfloor + 1}^{\min(N, s \lfloor QN \rfloor)}
P \{ W_{s, k} \}, \\
P_2(N, Q, s, \tau) &= \sum_{k = \lfloor QN \rfloor}^{\min(\lfloor \tau N \rfloor, (s+1) \lfloor QN \rfloor)}
P \{ W_{s+1, k} \}.
\end{split}
\eeq
The logarithmic asymptotics of the probability $P\{W_{s, k}\}$ was calculated
in~\cite{d15}, it equals
\beq{probWskAsym}
\lim_{N \to \infty} \frac{- \log_2 P \left\{ W_{s, \lfloor qN \rfloor} \right\} }{N} = \A(s, Q, q),
\eeq
where the function $\A(s, Q, q)$ is defined by~\req{exprA}.
Note that $P_1(N, Q, s, \tau) = 0$ if $\tau > sQ$ and $P_2(N, Q, s, \tau) = 0$ if $\tau < Q$.
This remark, \req{prob1Form} and \req{probWskAsym} yield
\beq{prob1Asympt}
\begin{split}
\lim_{N \to \infty} \frac{- \log_2 P_1(N, Q, s, \tau)}{N} &= \min_{\max\{\tau, Q\} \le q \le 1} \A'(s, Q, q),\\
\lim_{N \to \infty} \frac{- \log_2 P_2(N, Q, s, \tau)}{N} &= \min_{0 \le q \le \min\{\tau, (s+1)Q\}} \A'(s+1, Q, q),
\end{split}
\eeq
where the function $\A(s, Q, q)$ is defined by~\req{exprA'}.

Therefore, \req{prob1Asympt} and \req{probWskAsym} lead to the lower bound:
\beq{lowerRFirst}
R^{\tau}(s) \ge \max_{0 < Q < 1} \min \left\{ \min_{\max\{\tau, Q\} \le q \le 1} \frac{\A'(s, Q, q)}{s-1},
\min_{0 \le q \le \min\{\tau, (s+1)Q\}} \frac{\A'(s+1, Q, q)}{s} \right\}.
\eeq
Let us recall some analytical properties of the function $\A(s, Q, q)$~\cite{d15}. It is clear that
the function $\A(s, Q, q)$ as a function of the parameter $q$ decreases in the interval
$q \in [Q, 1 - (1 - Q)^s]$, increases in the interval $q \in [1 - (1 - Q)^s, \min\{1, sQ\}]$
and equals $0$ at the point $q = 1 - (1 - Q)^s$. Hence,
\begin{align*}
\min_{\max\{\tau, Q\} \le q \le 1} \frac{\A'(s, Q, q)}{s-1} &= 0 \quad \text{ if } \quad \tau \le 1 - (1 - Q)^s,\\
\min_{0 \le q \le \min\{\tau, (s+1)Q\}} \frac{\A'(s+1, Q, q)}{s} &= 0 \quad \text{ if } \quad \tau \ge (1 - Q)^{s + 1}.
\end{align*}
It means the equivalence of the bound~\req{lowerRFirst}
and the bound~\req{lowerRExpr}.
$\quad \square$

\textbf{Theorem 2.}\quad
\textit{As $s \to \infty$, the asymptotics of the optimal rate of threshold disjunctive codes
satisfies the inequality:}
\beq{lowerRAsympt}
R_{\text{Thr}}(s) \ge \frac{\log_2 e}{4s^3} (1 + o(1)), \quad s \to \infty.
\eeq

\textbf{Proof.}\quad
Our aim is to offer the lower bound for the asymptotic behaviour of the expression
\beq{fullExpr}
\max_{0 < \tau < 1} \; \max_{1 - (1 - \tau)^{1 / (s + 1)} < Q < 1 - (1 - \tau)^{1 / s}}
\min \left\{ \frac{\A'(s, Q, \tau)}{s - 1}, \frac{\A(s+1, Q, \tau)}{s} \right\},
\eeq
as $s \to \infty$.

For any fixed $\tau$, $0 < \tau < 1$, and any fixed $Q$, $1 - (1 - \tau)^{1 / (s + 1)} < Q < 1 - (1 - \tau)^{1 / s}$,
let us denote the solutions of the equation~\req{qSmall} for $\A(s, Q, \tau)$ and $\A(s+1, Q, \tau)$ by
$y_1(Q, \tau)$ and $y_2(Q, \tau)$. Note that $y_1$ can be greater than $1$.
It follows from~\req{qSmall} that the parameter $\tau$ can be expressed in the two forms:
$$
\tau = Q \frac{1 - y_1^s}{1 - y_1} = Q \frac{1 - y_2^{s+1}}{1 - y_2}.
$$
That is why the inequality $1 - (1 - \tau)^{1 / (s + 1)} < Q \Leftrightarrow \tau < 1 - (1 - Q)^{s + 1}$ is equivalent to
$$
\frac{1 - y_2^{s+1}}{1 - y_2} < \frac{1 - (1 - Q)^{s+1}}{1 - (1 - Q)},
$$
where, for any integer $n \ge 2$, the function $f(x) = \frac{1 - x^n}{1 - x}$ increases in the interval $x \in (0, +\infty)$.
Hence, we have
\begin{align*}
1 - (1 - \tau)^{1 / (s + 1)} < Q \quad &\Leftrightarrow \quad Q < 1 - y_2,\\
Q < 1 - (1 - \tau)^{1 / s} \quad &\Leftrightarrow \quad Q > 1 - y_1.
\end{align*}

In conclusion, the pair of parameters $(y_1, Q)$, $y_1 > 0$, $0 < Q < 1$, uniquely defines the parameters
$\tau$ and $y_2$. Moreover, if the inequalities
\beq{regionDef}
0 < \tau < 1, \quad
Q < 1 - y_2, \quad
Q > 1 - y_1.
\eeq
hold, then the parameters $\tau$ and $Q$ are in the region, in which the maximum~\req{fullExpr} is searched.

Let some constant $c > 0$ be fixed, $s \to \infty$ and $y_1 \eq 1 - c / s^2 + o(1 / s^3)$.
Then, the asymptotic behaviour of $\tau / Q$ equals
$$
\frac{1 - y_2^{s+1}}{1 - y_2} = \frac{\tau}{Q} = \frac{1 - y_1^s}{1 - y_1} = s - \frac{c}{2} + o(1),
$$
and, therefore,
$$
y_2 = 1 - \frac{c+2}{(s+1)^2} + o\(\frac{1}{s^3}\) = 1 - \frac{c + 2}{s^2} + \frac{2}{s^3} + o\(\frac{1}{s^3}\).
$$
To satisfy the inequalities~\req{regionDef} the parameter $Q$ should be in the interval
$$
\frac{c}{s^2} + o\(\frac{1}{s^3}\) = 1 - y_1 < Q < 1 - y_2 = \frac{c+2}{s^2} - \frac{2}{s^3} + o\(\frac{1}{s^3}\).
$$
Let us define the parameter $Q$ as $Q \eq d / s^2$, where $d$, $c < d < c + 2$, is some constant, and,
hence, $Q$ satisfies the previous inequalities.

The full list of the asymptotic behaviours of the parameters is presented below:
\beq{allAsympt}
\begin{split}
&\tau = \frac{d}{s} - \frac{cd}{2s^2} + o\(\frac{1}{s^2}\),\\
&Q = \frac{d}{s^2},\\
&y_1 = 1 - \frac{c}{s^2} + o\(\frac{1}{s^2}\),\\
&y_2 = 1 - \frac{c+2}{s^2} + o\(\frac{1}{s^2}\), \quad s \to \infty,
\end{split}
\eeq
where $c > 0$ is an arbitrary constant and $d = c + 1$.
The parameters defined by~\req{allAsympt} satisfy the inequalities~\req{regionDef}, and, therefore,
the substitution of asymptotic behaviours~\req{allAsympt} into~\req{fullExpr} leads to
some lower bound on the rate $R_{\text{Thr}}(s)$.

Let us calculate the asymptotics of
$$
\frac{\A(s, Q, \tau)}{\log_2 e} = (1 - \tau) \ln (1 - \tau) + (sQ - \tau) \ln \[\frac{1 - y_1}{Q}\] + s (\tau - Q) \ln y_1 - s(1 - Q) \ln (1 - Q).
$$
The first two terms of asymptotic expansion of the summands equals
\begin{align*}
(1 - \tau) \log_2 (1 - \tau) &= -\frac{d}{s} + \frac{cd}{2s^2} + \frac{d^2}{2s^2} + o\(\frac{1}{s^2}\),\\
(sQ - \tau) \ln \[\frac{1 - y_1}{Q}\] &= \frac{cd}{2s^2} \ln \[\frac{c}{d}\] + o\(\frac{1}{s^2}\),\\
s (\tau - Q) \ln y_1 &= -\frac{cd}{s^2} + o\(\frac{1}{s^2}\),\\
s(1 - Q) \ln (1 - Q) &= \frac{d}{s} + o\(\frac{1}{s^2}\).
\end{align*}
Therefore,
$$
\frac{\A(s, Q, \tau)}{\log_2 e} = \frac{d(d - c + c \ln [c / d])}{2s^2} + o\(\frac{1}{s^2}\)
$$

Further, let us calculate the asymptotics of
\begin{align*}
\frac{\A(s + 1, Q, \tau)}{\log_2 e} &= (1 - \tau) \ln (1 - \tau) + (sQ - \tau) \ln \[\frac{1 - y_2}{Q}\] + s (\tau - Q) \ln y_2 - s(1 - Q) \ln (1 - Q)\\
&+ Q \ln \[\frac{1 - y_2}{Q}\] + (\tau - Q) \ln y_2 - (1 - Q) \ln (1 - Q).
\end{align*}
The first two terms of asymptotic expansion of the new summands equals
\begin{align*}
(sQ - \tau) \ln \[\frac{1 - y_2}{Q}\] &= \frac{cd}{2s^2} \ln \[\frac{c+2}{d}\] + o\(\frac{1}{s^2}\),\\
s (\tau - Q) \ln y_2 &= -\frac{(c+2)d}{s^2} + o\(\frac{1}{s^2}\),\\
Q \ln \[\frac{1 - y_2}{Q}\] &= \frac{d}{s^2} \ln \[\frac{c+2}{d}\] + o\(\frac{1}{s^2}\),\\
(\tau - Q) \ln y_2 &= o\(\frac{1}{s^2}\).
\end{align*}
Therefore,
$$
\frac{\A(s+1, Q, \tau)}{\log_2 e} = \frac{d(d - c - 2 + (c + 2) \ln [(c + 2) / d])}{2s^2} + o\(\frac{1}{s^2}\)
$$

The maximum value of
$$
\max_{c > 0} \; \max_{c < d < c + 2} \; \min \left\{ d\(d - c + c \ln \[\frac{c}{d}\]\), d\(d - c - 2 + (c + 2) \ln \[\frac{c + 2}{d}\]\) \right\}
$$
is equal to $\frac{1}{2}$ and attained at $c \to \infty$ and $d = c + 1$.
$\quad \square$

\section{Upper Bounds}
\quad
As already mentioned in Remark~2 the paper~\cite{b15} considers the family of disjunctive $s$-codes
with a weaker condition than the definition of threshold disjunctive $s^T$-codes. These codes given as

\textbf{Definition 3.}~\cite{b15}.\quad
A binary code $X$ of length $N$ and size $t$ is said to be a {\em disjunctive $s_{\le T}$-code}
if the disjunctive sum of any $s$ codewords of $X$ has weight $\le T$ and covers those and only those
codewords of $X$ which are the terms of the given disjunctive sum.

Analogically, denote by $t'(N, s, L, T)$ the maximal size of disjunctive $s_{\le T}$-codes of length $N$,
introduce a parameter $\tau$, $0 < \tau < 1$, and define the \textit{rate} of disjunctive $s_{\le \lfloor \tau N \rfloor}$-codes as
\beq{RsLtau}
R_{\le \tau}(s) \eq \varlimsup_{N \to \infty} \frac{\log_2 t'(N, s, L, \lfloor \tau N \rfloor)}{N}.
\eeq

The paper~\cite{b15} provides the following upper bounds on the rate $R_{\le \tau}(s)$:
\beq{bonisBound}
R_{\le \tau}(s) \le \frac{\tau}{\lfloor s / 2 \rfloor^2 + \lfloor s / 2 \rfloor} \log_2 \frac{e s (s+2)}{4 \tau},
\eeq
which are also the upper bound on the rate $R^{\tau}(s)$.
Unfortunately, it gives inessential upper bounds on the optimal rate $R_{\text{Thr}}(s)$:
$$
R_{\text{Thr}}(s) \le \frac{8 \log_2 s}{s^2} (1 + o(1)), \quad s \to \infty.
$$
Nevertheless, if, for some constant $c > 0$, we substitute $\tau = c / s$
(the optimal value for constructions in Section~2 and for random coding bounds in Section~3)
into the lower bound~\req{bonisBound}, then we get $\frac{4 c \log_2 s}{s^3} (1 + o(1))$ as $s \to \infty$.
It is an interesting and open problem to obtain nontrivial upper bounds on the rate
of threshold disjunctive $s$-codes, we have a reason to suggest

\textbf{Hypothesis.}\quad
\textit{The rate of threshold disjunctive $s$-codes satisfies the inequality:}
$$
R_{\text{Thr}}(s) \le \frac{\text{Const} \cdot \log_2 s}{s^3} (1 + o(1)), \quad s \to \infty.
$$

\end{document}